\documentclass[preprint,review]{elsarticle}
\usepackage{epsfig}
\usepackage{graphicx}

\journal{Physica E}

\begin{document}

\begin{frontmatter}
  \title{Dynamic regime of electron transport in correlated
    one-dimensional conductor with defect} 
  \author[ire,mipt]{S N Artemenko}

  \ead{art@cplire.ru}

  \author[ire,mipt]{P P Aseev\corref{cor1}}
  \ead{pavel.aseev@phystech.edu} \author[ire]{D S Shapiro}
  \author[ire,mipt]{R R Vakhitov} \cortext[cor1]{Corresponding author}

  \address[ire]{Kotel'nikov Institute of Radio-engineering and
    Electronics of Russian Academy of Sciences, Moscow 125009, Russia}
%
%
  \address[mipt]{Moscow Institute of Physics and Technology,
    Dolgoprudny 141700, Moscow region, Russia}

\begin{abstract}
  The electron transport in a 1D conductor with an isolated local
  defect such as an impurity or a non-adiabatic contact is studied
  theoretically. New regime of conduction in correlated 1D systems is
  predicted beyond the well-known regime of tunneling resulting in the
  power-law I-V-curves. In this regime a quantum wire becomes
  ''opened'' at a voltage bias above the threshold value determined by
  $2k_F$-component of impurity potential renormalized by fluctuations,
  giving rise to a rapid increase of the dc current, $\bar I$,
  accompanied by ac oscillations of frequency $f = \bar
  I/e$. Manifestations of the effect resemble the Coulomb blockade and
  the Josephson effect. The spin bias applied to the system affects
  the I-V curves due to violation of the spin-charge separation at the
  defect site. The 1D conductor is described in terms of the
  Tomonaga-Luttinger Hamiltonian with short range or long-range
  Coulomb interaction by means of the bosonization technique. We
  derive boundary conditions that take into account relaxation in the
  leads and permit to solve non-equilibrium problems. Charge
  fluctuations are studied by means of Gaussian model which can be
  justified strictly in the limit of large voltages or strong
  inter-electronic repulsion. Spin fluctuations are taken into account
  strictly by means of the refermionization technique applicable in
  case of spin-rotation invariant interaction.
\end{abstract}


\end{frontmatter}
\section{INTRODUCTION}

It is well-known that in 1D systems the interaction between electrons
cannot be considered as a small perturbation and the system is
described as the Luttinger liquid (LL) that is an alternative to the
Fermi liquid for 1D electronic systems (for a review see
Ref.~\cite{Giamarchi,Voit}), and the Landau's Fermi-liquid picture
where low-energy excitations are single-electron quasiparticles that
in many respects behave like non-interacting electrons is not
applicable.  There are different realizations of 1D electronic systems
demonstrating properties of the LL. The examples are
semiconductor-based quantum wires in which dimensionality of the
conduction electrons is reduced by dimensional quantization
and 
carbon nanotubes, and such distinctive features of the LL as power-law
suppression of tunneling into 1D systems and spin-charge separation
and have been confirmed experimentally, see
e. g. Ref.~\cite{Cambridge}.

Electron-electron interaction greatly affects electronic transport in
1D systems. In particular, the back-scattering component of the
impurity potential in 1D systems with repulsive inter-electronic
interaction scales to infinity under renormalization group
transformations. Hence, even isolated impurities form effectively
large barriers and strongly suppress
conductance~\cite{KaneFisher,MatveevGlazman,FuruNag}.

On the other hand, the limit of strong interaction between electrons
in solids usually leads to the Wigner crystallization. However, in 1D
systems the long-range order is destroyed by
fluctuations~\cite{LL}. 
So, strictly speaking, 1D Wigner crystals do not exist, but the
density-density correlation functions of 1D gas with Coulomb repulsion
contain the $4k_F$ oscillating part which decays extremely
slowly~\cite{Schulz}, like $e^{-c\sqrt{\ln x}}$, that is slower than
any power-law. As the period corresponding to $4k_F$ oscillations is
exactly the average inter-electron spacing, such a system can be
considered as a 1D Wigner crystal with pseudo-long-range
order~\cite{Schulz}. In case of short range inter-electronic
interaction (which takes place in gated quantum wires where the
long-range part of the Coulomb interaction is screened by electrons in
the metallic gate) the $4k_F$ density correlations decay slowly as
well, as the power-law with a small exponent.

Sliding of electronic crystals contributes to conductance, the most
studied case being quasi-1D CDW compounds~\cite{Gruner}. Defects pin
the CDW but when the driving electric field exceeds a threshold field
the CDW starts to slide resulting in non-linear conductance and ac
generation at washboard frequencies corresponding to a shift of the
CDW by one period~\cite{Gruner}. As long as the LL can be interpreted
as a 1D form of the 1D Wigner crystal, one can expect a similar
dynamic regime of depinning, sliding and ac generation in correlated
1D electron system as well.  We show that such a regime does exist, at
least, in the quasiclassical limit when quantum fluctuations at the
impurity site are suppressed by strong electron-electron interaction.
Such a scenario was addressed earlier in our letter~\cite{ARS} where
the dynamic regime of conduction accompanied by oscillations of
frequency $f = \bar I/e$ was predicted in a spinless LL.

Full I-V curves of a single-channel LL with a single impurity were
studied by means of thermodynamic Bethe ansatz technique by Fendley et
al~\cite{FeLuSa}. Egger and Grabert~\cite{Grabert} calculated the I-V
curves for specific value of interaction parameter $K_\rho =1/2$ using
the refermionization technique which makes the Hamiltonian quadratic
and, hence, solvable exactly. But no non-stationary regime was found.
Possibility of generation of self-sustained current oscillations in a
quantum wire in a properly designed load circuit was considered in
Ref.~\cite{EgKoutouza}, but these oscillations are a consequence of
instability induced by S-shaped I-V curves, and their origin is
different from the mechanism discussed in the present work.  We
suppose that the main difference between our approach and
Refs.~\cite{FeLuSa,Grabert,EgKoutouza} is that the equilibrium
distribution of incident particles (non-interacting fermions, kinks
and anti-kinks, etc) was assumed in these papers. However, as the
distribution of the particles transmitted through the defect is not
the equilibrium one, and the bosonic excitations of the LL are
reflected from the leads to the quantum wire even in case of adiabatic
contacts since the reflection coefficient $r =
\frac{1-K_\rho}{1+K_\rho}$~\cite{SafiSchulz}. Then the incident waves
consist in part of the particles reflected from the contact. So if the
relaxation inside the conducting channel is small the distribution of
the incident particles must not be necessarily the equilibrium one,
and this applies equally to fermions derived from bosons after the
refermionization. Therefore, one needs to calculate the distribution
function of the incident particles, and we perform this by means of
boundary conditions which take into account relaxation processes
induced by coupling of the quantum wire to the Fermi liquid of the
current leads considered as a heat bath. These boundary conditions are
valid for non-ideal contacts, and they generalize the boundary
conditions by Egger and Grabert~\cite{Grabert} and the results of Safi
and Schulz~\cite{SafiSchulz,Safi} derived for expectation values and
ideal adiabatic contacts.

We think that the results of Refs.~\cite{FeLuSa,Grabert,EgKoutouza}
are applicable in the limit of conducting channels longer than the
damping length of excitations due to coupling of electrons inside the
wire to a dissipative bosonic bath (phonons, density fluctuations in a
metallic gate, and so on). And we obtain the non-stationary regime of
conduction for practically important case of the quantum wire which is
shorter than the relaxation length, so that the relaxation is governed
by boundary conditions.

The structure of the paper is as follows. In Sec.~\ref{form} we
formulate the problem, derive boundary conditions at the contacts, and
derive equations of motion for the displacement field at the impurity
position. These equations resemble equations of motion of cou0led
quantum pendulums. In Sec.~\ref{spinless} we use our equations to
study electronic transport in spinless LL. Using the Gaussian model to
account for fluctuations, we study I-V curves, analyze noise spectrum,
study non-Gaussian corrections and find that the Gaussian
approximation is justified in the limit of strong interaction between
electrons and large voltages. In Sec.~\ref{spinful} we consider the
spinful LL with strong enough interaction between electrons when
charge fluctuations at the defect position are small. However, spin
fluctuations are large and they are taken into account strictly by
means of refermionization method in spin sector valid in case of
spin-rotation invariant interaction ($K_\sigma =1$). In
Sec.~\ref{N-ideal} we show that non-adiabatic contacts induce
non-stationary effects similar to those induced by impurities. In
Sec.~\ref{concl} we formulate conclusions.

Below we set $e$, $\hbar$ and $k_B$ to unity, restoring dimensional
units in final expressions when necessary.

\section{GENERAL FORMULATION}\label{form}
\subsection{Problem formulation}

We consider a correlated 1D conductor with an impurity at $x=0$ and
connected to ideal Fermi-liquid reservoirs at $x= \pm L/2$. The
Hamiltonian of the system with impurity consists of two terms $ H=H_0+
H_{i}.  $ The first one is the bosonised Tomonaga-Luttinger (TL)
Hamiltonian that maps the 1D system of interacting electrons to free
massless bosons described in terms of the displacement fields
$\hat\Phi_{\nu} (t,x)$ and the conjugated momentum density
$\hat\Pi_\nu (t,x) = \partial_x \hat\Theta_{\nu}/\pi$.  Here $\nu =
\rho, \sigma$ denotes charge and spin channels, correspondingly.  The
standard TL Hamiltonian in the Fourier transformed form
reads~\cite{Giamarchi,Voit}
\begin{equation}
  \hat H_0 = \frac{\pi v_{F}}{2} \sum_{\nu=\rho,\sigma} \int
  \frac{dq}{2 \pi} \left\{\hat \Pi_\nu^2 +
    \frac{1}{\pi^2 K_\nu^2} q^2 \hat \Phi_\nu ^2\right\}. \label{H0}
\end{equation}
Here the LL parameters $K_\nu$, playing the role of the stiffness
coefficients of the elastic string described by Hamiltonian
(\ref{H0}), are related to the electron-electron interaction
potential, and measure the strength of interaction between
electrons. In the spin-rotation invariant case considered in our
study, $K_\sigma =1$, $K_\rho(q)= 1/\sqrt{1+\frac{g(q)}{\pi v_F}}, $
where $g(q)$ is the Fourier transformed interaction potential. In case
of the short-range interaction the dependence of $g$ on wave-vector
$q$ is usually neglected. For repulsive interaction $K_\rho <1$.  In
infinite 1D gas with long-range Coulomb interaction described by the
approximate form $V_C(x) = \frac{e^2}{\epsilon \sqrt{x^2 +d^2}}$,
where $\epsilon$ is a background dielectric constant and $d$ is a
diameter of quantum wire, one obtains $g(q) = 2 \frac{e^2}{\epsilon}
{\rm K_0} (|qd|)]$~\cite{Schulz}. Thus,
\begin{equation}
  K_\rho(q)=\frac{1}{\sqrt{1+\gamma  {\rm K_0} (|qd|)}}, \quad \gamma = \frac{2 e^2}{\pi \hbar v_F \epsilon}  \approx  \frac{2}{137\pi} \left(\frac{c}{v_F }\right) \frac{1}{\epsilon} , \label{gamma}
\end{equation}
where $\gamma$ is dimensionless parameter which measures the strength
of the Coulomb repulsion between the electrons.

In case of the long-range interaction and finite length of the
conducting channel the Coulomb potential is modified by screening of
the interaction by current leads. The exact form of the screening
depends on the geometry of the system. We consider 3D metallic leads
forming sheets of a plane capacitor connected by the quantum
wire. Then the screening by the leads can be depicted in terms of the
image charges, and the interaction potential between charges located
at $x$ and $x'$ is described as
\begin{equation}
  V(x,x') =  \sum_{n=-\infty}^{\infty} \left[V_C (x-x'+2nL) - V_C (x+x'+2nL+L)  \right],
  \label{pot}
\end{equation}
where the term with $n=0$ describes the direct Coulomb interaction,
and other terms are induced by image charges.  Its contribution to the
$\nu = \rho$ term in the Hamiltonian (\ref{H0}) in the coordinate
representation reads
\begin{equation}
  \int dx dx'
  \left\{\partial_x \hat\Phi_\rho (x) V(x,x')
    \partial_{x'} \hat\Phi_\rho (x') \right\}
\end{equation}
Since the operator of the particle density is given by expression
$\hat\rho = -(\sqrt{2}/\pi)\partial_x\hat\Phi(x)$, this term has
rather transparent physical meaning.

Interaction with the impurity is described in terms of the phase
fields $\hat\Phi_\nu (t,x)$ at the impurity position $x=0$
\cite{Giamarchi,Voit}
\begin{equation}
  \hat H_{i} = - \frac{W}{\pi}   \cos{\sqrt{2} \hat\Phi_\rho (0)}\cos{\sqrt{2} \hat\Phi_\sigma (0)},
  \label{impu}
\end{equation}
where the impurity strength $W$ is related to the back-scattering part
of the impurity potential. The forward scattering is not included
because it can be eliminated from the problem by redefinition of the
field $\hat\Phi_\rho$~\cite{Giamarchi}. The impurity Hamiltonian is
related to $2k_F$-components of electron density and in the Luttinger
model used here it does not contain higher harmonics, which are
present in more general models~\cite{Voit}.

Current in the system can be calculated in terms of $\hat\Phi_\rho$ by
means of thermodynamic averaging of the expressions for the operator
\begin{equation}
  \hat I = \frac{\sqrt{2}}{\pi}\partial_t \hat \Phi_\rho. \label{i}
\end{equation}
The expectation value of the displacement field in (\ref{i}) can be
found from equation of motion for the Heisenberg operator
$\hat\Phi_\rho (t,x)$. Commuting $\hat\Phi_\rho$ with the Hamiltonian
we find for the case of short range interaction
\begin{equation}
  \left(  v_\rho^2 \partial^2_{x} - \partial^2_{t}\right)\hat\Phi_\rho (t,x) = \sqrt{2} \pi v_F  W  \sin \sqrt{2} \hat\Phi_{\rho } \cos{\sqrt{2} \Phi_\sigma}  \delta (x) ,
  \label{phiop}
\end{equation}
where $v_\rho = v_F/K_{\rho}$ is the velocity of charge (plasmonic)
excitations. Equation of motion for the spin field has similar form,
it can be obtained from (\ref{phiop}) by substitution subscripts
$\rho$ by $\sigma$ and vice versa.

At the contacts we apply the boundary conditions which take into
account injection of electrons induced by external bias and relaxation
processes induced by coupling of the quantum wire to 2D or 3D Fermi
liquid in the current leads. The boundary conditions are considered in
details in the next subsection.

\subsection{Boundary conditions}

Boundary conditions for the single mode (spinless or spin-polarized)
wire contacting with a 2D or 3D leads were derived in
Ref.~\cite{ArtAsSh}. Here we generalize this result for the spinful
case. In order to derive the boundary conditions we use the ideas of
the scattering approach (for a review see Ref.~\cite{Buttiker}).

We assume that electrons in the leads do not interact and that
longitudinal (along the $x$-axis) and transverse motions are
separable. Here we concentrate on the case of 
contacts at $x=\pm L/2$ with an arbitrary transverse profile of the
potential. The longitudinal motion in the leads is characterized by
wave vector $k$, spin $s$ and energy $\varepsilon_l =
\frac{k^2}{2m}$. The transverse motion is described by energy
$\varepsilon_n$, the total energy being $\varepsilon = \varepsilon_l +
\varepsilon_n$, where $n$ is an index labeling transverse modes.

In case of non-interacting electrons we match the electron field
operators in the lead and in the wire and, using independency of the
annihilation operators $\hat c_{n,k}$ of the incident  electrons on the properties of the
contact, we derive boundary conditions for the lowest subband, which
is responsible for an electronic transport in the wire. The detailed derivation is given in the Appendix.

It is convenient to express the boundary conditions in terms of physical
values: the current $\hat j = v_F\left(\hat \psi_R^\dag \hat \psi_R -
  \hat \psi_L\hat \psi_L \right)$, the smooth part of charge density
perturbations $\hat \rho = \hat\psi_R^\dag \hat\psi_R +
\hat\psi_L^\dag\hat\psi_L$, and the $2k_F$-component of charge density
perturbations $\hat \rho_F = \hat \psi^\dag_{L} \hat \psi_R e^{2iq_F
  x} + c.c.$, which is related to the Friedel oscillations, where
$\hat \psi_{R,L}$ are field operators for right and left moving
electrons in the wire. The details of derivation can be found in the
appendix. Then the boundary conditions at the left(right) contact read
\begin{equation}
  \frac{v_F}{T}\hat \rho \pm \hat j + v_F f \hat \rho_F  =
  \frac{1}{V} \sum_{\mathbf{n},\mathbf{n'}}\hat c_{\mathbf{n'}}^+ \hat c_{\mathbf{n}} e^{i(\varepsilon_{\mathbf{n'}}-\varepsilon_{\mathbf{n}})t}.
  \label{bc-rho}
\end{equation}
Here $T$ is a parameter that
characterizes reflection from the contact, and $T=1$ corresponds to an
adiabatic contact. Parameter $f$ descibes the amplitude of the Friedel
oscillations, it  is a number of the order unity if $T$ is not close to
unity, and $f \simeq \sqrt{2(1-T)}$ if the contact is neally adiabatic, $1-T\ll 1$. Thus the Friedel
oscillations disappear if the contacts are ideal. These parameters are local in the
sense that they depend only on the properties of the given contact and
do not depend neither on the lead at the opposite end of the 1D channel nor on the presence
of an impurity or electron-electron interaction provided the latter vanishes in the leads. The
explicit expressions for $T$ and $f$ are given in the Appendix.

In order to check the validity of conditions~(\ref{bc-rho}), we
considered a wire with non-interacting 1D electrons attached to
smoothly widening nearly adiabatic leads. We also assumed that there
might be a potential step of the height $U_0 \ll \varepsilon_F$ at the
interface. In this case we can find the solution directly, using the
quasiclassical approximation in the lead and matching the
quasiclassical solution outside the 1D conductor with the exact
solution inside the channel. And we found that the
condition~(\ref{bc-rho}) fulfils again and yields the conductance
$G=TG_0$ in agreement with the Landauer formula.

As we need the boundary conditions in the bosonic representation, we
have to bosonize (\ref{bc-rho}). Note that the LL theory is valid
provided that all energies are small in comparison with the Fermi
energy, while the amplitude of the term $v_F f \hat \rho_F$ which is
responsible for the Friedel oscillations is of the order of the Fermi
energy if $f =2\sqrt{1-T}$ is not small. Therefore, we limit our study
by nearly adiabatic contacts with $\sqrt{1-T} \ll 1$, and neglect
terms of the higher order in $f$. Transforming then in a standard way
the fermionic operators to charge and spin density
variables~\cite{Voit} we obtain the boundary conditions for bosonic
field $\hat \Phi_\rho$ at the left (right) contacts
\begin{eqnarray}
  &&
  v_F \partial_x \hat\Phi_\rho \mp \partial_t \hat\Phi_\rho + \sqrt{2}f\varepsilon_F \sin (\sqrt{2}\hat\Phi_\rho \mp k_FL)\cos\sqrt{2}\hat\Phi_\sigma = \hat P^{L,R}_\rho, \label{bc-operator1} \\
  &&
  v_F \partial_x \hat\Phi_\sigma \mp \partial_t
  \hat\Phi_\sigma + 
  \sqrt{2}f\varepsilon_F \cos (\sqrt{2}\hat\Phi_\rho \mp k_FL) \sin \sqrt{2} \hat\Phi_\sigma
  = \hat P^{L,R}_\sigma , \label{bc-operator-sigma}
\end{eqnarray}
where $\hat P_\nu^{L,R} = 2\pi v_F \hat N_\nu^{L,R}$, $N_\nu^{L,R}$ is
the operators of excess number of charge ($\nu=\rho$) and spin
($\nu=\sigma$) densities in the left ($L$) and right ($R$) leads,
respectively. The expectation values of the operators $P^{L,R}_\nu$
and correlation functions of their fluctuating parts $\delta\hat
P^{L,R}_\nu = \hat P^{L,R}_\nu - \langle \hat P^{L,R}_\nu \rangle$ can
be calculated easily from the right-hand part of (\ref{bc-rho}). The
average of $P^{L,R}_\rho$ for charge channel is proportional to the
potentials $U^{L,R}_\rho$ applied to the left (right) contact,
$\langle P^{L,R}_\rho \rangle = U^{L,R}_\rho/\sqrt{2}$. Similarly, the
expectation values $\langle \hat N_\sigma^{L,R}\rangle$ equal to the
excess spin densities in the leads, and $\langle \hat P_\sigma^{R} -
\hat P_\sigma^{L} \rangle = V_\sigma/\sqrt{2}$ where $V_\sigma$ is a
``spin bias''.

Correlation functions are identical for both channels and for both
contacts, while correlations between left and right contacts and
between charge and spin operators are absent. In the frequency
representation correlation functions read
\begin{equation}
  \langle \delta\hat P (\omega)\delta\hat P (\omega') \rangle = 4\pi^2  \omega N(\omega') \delta (\omega + \omega').
  \label{P-corr}
\end{equation}
where $N(\omega')$ is the Planck distribution function. The
fluctuating part of the boundary conditions takes into account that
the leads play a role of a heat bath and leads to the equilibrium
distribution functions of the excitations in the quantum wire.

If there is a metallic gate near the quantum wire we must take into
account screening by the gate. Following the approach of
Ref.~\cite{Grabert} we find that the screening by the gate results in
a modification of the factor in the first term of
(\ref{bc-operator1}). Then the boundary conditions for the case of
short-range interaction acquire the form
\begin{equation}
  \frac{v_F}{K_\rho^2} \partial_x \hat\Phi_\rho \mp \partial_t \hat\Phi_\rho + \sqrt{2}f\varepsilon_F \sin (\sqrt{2}\hat\Phi_\rho \mp k_FL)\cos\sqrt{2}\hat\Phi_\sigma  = \hat P^{L,R}_\rho  . \label{bc-operator-rho}\\
\end{equation}

The modification of the factor before the spatial derivative can be
also illustrated by means of the simple model in which the factor
$K_\rho$ is equal to $1$ at the non-interacting lead $x=-L/2-0$ and
step-like reaches its value in the wire at $x=-L/2+0$. Then we
integrate the equation of motion~(\ref{phiop}) from $x=-L/2-0$ to
$x=-L2/+0$ and obtain that $\hat \Phi_\rho$ is a continuous function
of $x$ but its spatial derivative satisfies
$$
\partial_x \hat \Phi_\rho(-L/2-0) = \frac{1}{K_\rho^2}\partial_x \hat
\Phi_\rho(-L/2+0),
$$
which explain transition
from~(\ref{bc-operator1})~to~(\ref{bc-operator-rho})

In case of a wire adiabatically connected to ideal Fermi-liquid
reservoirs at $x= \pm L/2$ the boundary conditions
(\ref{bc-operator-sigma}) and (\ref{bc-operator-rho}) reduce to
\begin{equation}
  \left ( \frac{v_F}{K_{\nu}^2}\partial_x \mp \partial_t \right )\hat\Phi_\nu (x{=} \pm L/2){=}\hat P_\nu^{L,R}  , \qquad \nu = \rho, \sigma  \label{bc-operator}
\end{equation}
in agreement with the results of Ref.~\cite{Grabert,SafiSchulz,Safi}.

It looks natural that in case of gated quantum wire the gate screens
externally applied electric field and the problem is described in
terms of boundary conditions, as it was discussed in
Ref.~\cite{Grabert}. Of course, inside the wire there is also an
electric field induced by non-uniform distribution of electrons, but
this electric field is taken into account by the interaction between
electrons. However, it looks less clear whether one can describe the
driving voltage by boundary conditions when there is no gate (the case
of long-range interaction). Therefore, in case of long-range Coulomb
interaction we considered two approaches. First, we inserted the
driving dc electric field into the Hamiltonian, when the external
field appears in the equation of motion for the displacement field
$\hat\Phi_\rho(x,t)$. Second, we derived equations of motion for the
phase fields with driving dc voltage taken into account by boundary
conditions. But the equation of motion for the displacement field
$\hat\Phi_\rho(t)$ at the defect site turned out to be the same and
the results of two approaches for the case of dc voltage in both cases
are equivalent.

\subsection{Equations of motion of the displacement field at the
  impurity site}\label{equations}

In this section we derive equations of motion for the phases
$\hat\Phi_{\rho }$ and $\hat \Phi_{\sigma}$ at the impurity for the
wire with adiabatic contacts. Consider first the case of short-range
interaction. We solve equation of motion (\ref{phiop}) for
$\hat\Phi_\nu (\omega,x)$ formally using Fourier transformation with
respect to time, and match the solutions at the impurity site using
boundary conditions (\ref{bc-operator}).  In this way we express
operators $\hat\Phi_\nu (\omega,x)$ in terms of their values at the
impurity site, $x=0$, and after inverse Fourier transformation obtain
equations of motion for the displacement field at the impurity
site. The equations read
\begin{eqnarray}
  &&
  \partial_t \hat \Phi_{\rho} +   \frac{W}{\sqrt{2}}   Z \otimes \sin \sqrt{2}\hat \Phi_{\rho}\cos \sqrt{2}\hat \Phi_{\sigma}  
  =  F \otimes  \hat P_\rho   ,
  \label{em-rho}
  \\
  &&
  \partial_t \hat \Phi_{\sigma} +   \frac{W}{\sqrt{2}} \sin \sqrt{2}\hat \Phi_{\sigma}\cos \sqrt{2}\hat \Phi_{\rho}  
  =  \hat P_\sigma \left(t-\frac{L}{2v_F} \right)  .
  \label{em-sigma}
\end{eqnarray}
Here $\otimes$ means convolution in time, $P_\nu = \hat P_\nu^{R} -
\hat P_\nu^{L}$, $Z(t)$ and $F(t)$ are defined by means of Fourier
components
\begin{equation}
  Z(\omega)= K_{\rho} \frac{1- iK_{\rho} \tan \omega t_L}{K_{\rho} -i \tan \omega t_L},  \quad F (\omega) =  \frac{K_{\rho} }{2[K_{\rho} \cos \omega t_L -i \sin \omega t_L] },
  \label{ZF}
\end{equation}
where $\quad t_L = \frac{L K_\rho}{2v_F}$.  Oscillatory dependence of
$Z(\omega)$ and $F(\omega)$ describes multiple reflections of the
bosonic excitations of the LL from contacts. This statement can be
illustrated by the expression for $Z$ in time representation
\begin{equation}
  Z(t)  =  K_\rho \left[\delta (t)+ 2 \sum_{m=1}^\infty r^m  \delta \left(t - m t_L \right) \right], \quad r=\frac{1-K_\rho}{1+K_\rho},
  \label{z}
\end{equation}
where $r$ is the reflection coefficient of plasma excitations from the
contacts~\cite{SafiSchulz}.

Consider now the case of long-range Coulomb interaction between the
electrons. Formally, the interaction potential in the system of finite
length (\ref{pot}) is symmetric with respect to the contacts and
periodic with period $2L$. Therefore, we can expand the field
operators in Fourier series and find a simple and easily soluble
equation of motion for Fourier components.  Then using the boundary
conditions we obtain equations of motion similar to
(\ref{em-rho}-\ref{em-sigma}) but with different memory functions $F$
and $Z$
$$
Z(\omega) = \frac{i\omega R_+ - 2\omega^2(R_+^2 - R_-^2)}{1+2i\omega
  R_+ },\, F(\omega) = \frac{i\omega R_- }{1+2i\omega R_+ }, \,
R_{\pm}(\omega) = \frac{v_F}{L}\sum_{k=-\infty}^{\infty} \frac{(\pm
  1)^k}{\omega^2 - q_{2k}^2 v^2(q_{2k})} $$ with $ q_n = \frac{\pi
  n}{L}$, $v^2=v_F^2[1+2\gamma {\rm {\rm K}_0} (|qd|)]$. The exact
analytical summation in $R_{\pm}$ is difficult, but the sums can be
calculated with logarithmic accuracy as
$$
R_{+}(\omega) = \frac{K_\rho(q_\omega)}{2\omega \tan \frac{\omega L}{2 v_\omega}}, \quad R_{-}(\omega) = \frac{K_\rho(q_\omega)}{2\omega \sin \frac{\omega L}{2 v_\omega}},\quad K_\rho(q_\omega) = \frac{1}{ \sqrt{1+2\gamma {\rm {\rm K}_0} (|q_\omega d|)} },
$$
where $q_\omega$ is a solution of equation $\omega= q_\omega v(q_\omega)$. This approximation results in expressions for $Z$ and $F$ that coincide with (\ref{ZF}) but with  $K_\rho(q_\omega)$ depending on frequency.

In the simplest case of the single-mode (spinless) LL the equation of motion for the phase at the impurity site reads
\begin{equation}
  \partial_t \hat \Phi (t) +    W_i   Z \otimes \sin 2 \hat\Phi  =   F\otimes\hat P.
  \label{phi0op} 
\end{equation}

Equations (\ref{em-rho}-\ref{em-sigma}) and (\ref{phi0op}) resemble equations of motion of an overdamped pendulums, therefore, one can expect that when the system is driven by a constant external bias the phase increases non-uniformly, which in our case means presence of both dc and ac current. It is not easy to solve the non-linear equations for operators in general case. So we solve them in the limit of strong inter-electronic interaction when fluctuations of the phase field $\hat \Phi_{\rho}$ are relatively small and can be described by Gaussian approximation. Fluctuations in the spin channel  are not small and are not Gaussian, however, they will be taken into account strictly by means of refermionization.

\section{DYNAMIC REGIME OF CONDUCTION IN THE SPINLESS LUTTINGER LIQUID}\label{spinless}

\subsection{Gaussian approximation}

In this section we will consider the most technically simple case of the single-mode LL with short-range interaction between electrons. 

First, we represent the bosonic field operator at the impurity site as a sum of its expectation value and fluctuating part, $\hat\Phi  = \Phi + \hat\phi$, $\Phi = \langle \hat\Phi\rangle$. Then we perform thermodynamic averaging of both sides of Eq.~(\ref{phi0op}) and obtain equation for expectation value $\Phi$ of the field operator at the impurity site
\begin{equation}
  \partial_t \Phi (t) +    W_i   Z \otimes \langle \sin 2 \hat\Phi \rangle =   F\otimes V ,
  \label{phi0}
\end{equation}
 
Equation (\ref{phi0}) is not a closed equation for $\Phi (t)$ since it contains an expectation value of $\sin 2 \hat\Phi (t)$ which depends both on expectation value $\Phi$ and on fluctuations $\hat\phi$ of the displacement field. Therefore, in order to calculate the expectation value we need to study fluctuations. The equation of motion for the fluctuating part $\hat\phi$ of the displacement field we obtain subtracting (\ref{phi0}) from (\ref{phi0op}). Then we simplify the problem assuming that fluctuations are Gaussian. Strictly speaking, the fluctuations are not Gaussian, and in general case this is just a model assumption. However, we show below that this approach can be justified in case of strong inter-electronic repulsion and in the limit of high voltages, where the Gaussian fluctuations dominate. 

Thus we solve the problem by means of the self-consistent harmonic approximation~\cite{Giamarchi}, in which fluctuations are assumed to be Gaussian. In this approximation, we replace  
\begin{equation}
  \sin 2 \hat\phi \to  2 h \hat\phi, \quad h \equiv e^{-2 \langle \hat\phi^2 \rangle} ,
  \label{scha}
\end{equation}
and instead of (\ref{phi0}) we obtain more simple equation for the expectation value $\Phi (t)$
\begin{equation}
  \partial_t \Phi (t) +    W_i   Z\otimes h   \sin 2 \Phi    =   F \otimes V ,
  \label{phi0s}
\end{equation}
and a linear equation for fluctuations
\begin{equation}
  \partial_t \hat\phi (t) +  2 W_i  Z \otimes h \cos2\Phi \hat\phi  =   F \otimes \delta\hat P(t_1) .
  \label{phi-flu}
\end{equation}
Coefficients of this equation depend both on the mean square fluctuations $\langle \hat\phi^2 (t) \rangle$ and on the expectation value $\Phi$, so it must be solved self-consistently with (\ref{phi0s}).

If the applied dc voltage is small enough, 
equations (\ref{phi0s}) and (\ref{phi-flu}) have stationary solutions for phase $\Phi$ and for mean square fluctuations $\langle \hat\phi^2  \rangle$. 
In the stationary case (\ref{phi0s}) reads
\begin{equation}
  W_i h \sin 2 \Phi  = V ,
  \label{phi-st}
\end{equation}
and Fourier transformed (\ref{phi-flu}) reduces to the simple form
\begin{equation}
  -i\omega \hat\phi (\omega) +  2 W_i h Z(\omega) \cos2\Phi \hat\phi (\omega) =   F(\omega) \delta\hat P(\omega) .
  \label{eqflu-st}
\end{equation}
This equation can be solved easily. Taking into account correlation functions given by (\ref{P-corr}), (\ref{eqflu-st}) and (\ref{ZF}) we can calculate mean square fluctuations 
\begin{equation}
  \langle \hat\phi^2 \rangle  =  \frac{K_\rho^2}{2} \int_{-\infty}^\infty  \frac{\omega \coth\frac{\omega}{2T} d\omega}{(\omega^2 + W_c^2)[(1+K_\rho^2) + (1- K_\rho^2)\sin (\omega t_L - \alpha_\omega)]},
  \label{flu-st}
\end{equation}
where
$
\alpha_\omega = \arctan \frac{W_c^2 - \omega^2}{2\omega W_c}, \quad W_c= 2W_i K_\rho h \cos 2\Phi.
$ 
Since $W_c$ depends on $\langle \hat\phi^2 \rangle$, (\ref{flu-st}) determines the self-consistency condition for $\langle \hat\phi^2 \rangle$. The result of integration depends on relation between $V_T$
and temperature $T$. First, we consider the limit of zero
temperature. In pure LL this integral would diverge
logarithmically both at high and low frequencies. The
divergence at the upper limit in the TL formalism must be
cut off at frequency $\Lambda$ of the order of the bandwidth
or the Fermi energy. The infrared divergence at low
frequencies is a distinctive feature of 1D
systems, and in the presence of impurity the infrared divergence is cut off at a frequency related to the impurity potential. 
In addition, the
denominator contains the oscillating factor induced by
reflections of fluctuations from contacts. If the length of
the quantum wire is large enough the main
contribution to the integral is determined by frequencies
$\omega t_L \gg 1$ and oscillations contribute little to the
integral and we obtain 
\begin{equation}
  \langle \hat\phi^2 \rangle = \frac{K_\rho}{2(1-K_\rho)} \ln \frac{\Lambda}{2 W_i K_\rho \cos 2\Phi}.
  \label{phi2}
\end{equation}
Now using this equation we can calculate maximum value
of the left hand side of (\ref{eqflu-st}) which
determines the value of the threshold voltage  $V_T$ below which
the static solutions for mean phase $\Phi$ exist. We find
\begin{equation}
  V_T = 2W_i \left( \frac{2W_i \sqrt{K_\rho^3} }{\Lambda}\right)^{\frac{K_\rho}{1-K_\rho}}\sqrt{1-K_\rho}.
  \label{VT}
\end{equation}
We see that the threshold voltage at low temperatures is
determined by the impurity potential renormalized by quantum
fluctuations. In case of interelectronic repulsion, $K_\rho
< 1$, the mean square fluctuations $\langle \hat\phi^2
\rangle$ and, hence, $V_T$ are finite, while in
non-interacting system, when $K_\rho = 1$, fluctuations
become infinite and $V_T$ is destroyed by quantum
fluctuations. Thus we find that the solution for $\Phi$ is
stationary at $V<V_T$, that is current cannot pass an
impurity. This result is a consequence of our approximation
in which only Gaussian fluctuations were taken into account.
If we took into account fluctuations of solitonic type for
which the phase increases by $\pi$ due to tunneling, we
would obtain a small tunneling current at $V < V_T$ described by the well-known power-law
I-V curves~\cite{Giamarchi}. Thus the current is small at $V < V_T$ and starts to increase rapidly at $V > V_T$.

In case of finite temperatures the
self-consistency equation has solutions which correspond to
a finite value of fluctuations only if $T < T_c \sim V_{T,0}
\equiv V_{T}(T=0, L=\infty)$, so there is a characteristic
temperature above which $V_T$ is destroyed by thermal
fluctuations and the impurity does not suppress electronic
transport.
 
If the quantum wire is short enough, $L \sim v_\rho/V_{T,0}$, we must not average (\ref{flu-st}) over oscillations at $\omega t_L \sim 1$. At these frequencies $\langle \hat\phi^2(\omega) \rangle$ in (\ref{flu-st}) is proportional to $\omega^{-1}$ as before but with a different factor. As a consequence $V_T$ is suppressed in short wires, and impurities do not destroy the linear conduction when $L < L_c \sim v/V_{T,0}$. This happens due to increase of fluctuations at the impurity site because fluctuations are reflected back from the contacts, while the distance to the contacts becomes smaller than the correlation length of the fluctuations. 

\subsection{I-V curves and noise spectrum at high voltages}

As it was noted already, it is difficult to obtain I-V curves at low voltages accurately because of time dependence of the mean square value of fluctuations. The problem is simplified at high voltages, $V \gg V_T$, when the mean square value $\langle \hat\phi^2 \rangle$ becomes nearly constant with small oscillating component. In this case (\ref{phi0s}-\ref{phi-flu}) can be solved perturbatively assuming that the oscillating parts of both mean square fluctuations $\langle \hat\phi^2 \rangle$ and of the mean phase $\Phi$ are small. 

In this subsection we consider the limit of relatively long
conducting channel, $V_T t_L \gg 1$, but not too long, so
that the wire is short in comparison with the damping length
related to relaxation due to coupling to phonons etc. In this case we have to use the exact form  of
$Z(t)$ in equation for the expectation value
(\ref{phi0s}) but can keep only the first
delta-function in kernel $Z(t)$ in equation for fluctuations
(\ref{phi-flu}). In time representation this means that
we take into account current pulses reflected from the
contacts but we ignore correlations between fluctuations
shifted by time $n t_L $ necessary for the excitation to
return to the impurity after multiple reflections
from the contacts. Then (\ref{phi-flu}) acquires simple
form and can be solved easily 
\begin{equation}\label{hp11}
  \hat \phi = \int_{-\infty}^{t} dt_1 \int_0^\infty dt_1  F(t_1-t_2) \delta\hat P(t_2)    e^{-\int_{t_1}^{t} W(t_2) dt_2 },  
\end{equation}
where $ W(t) = 2 K_\rho W_i h(t) \cos2\Phi (t)$. Using (\ref{hp11})  we can calculate mean square fluctuations $\langle \hat\phi^2 \rangle$. As we consider the long channel we average, again, over oscillatory factor in $F(t)$ and find
\begin{equation}
  \langle \hat\phi^2 \rangle = 
  \frac{K_\rho}{4}  \int_{-\infty}^{t}  dt_1 \, dt_{3} \int d\omega \omega \coth\frac{\omega}{2T} e^{-\int_{t_1}^{t} W(t_2)dt_2 - \int_{t_3}^{t} W(t_2) dt_2-i\omega(t_1-t_3)}.
  \label{cor}
\end{equation}
To solve this equation we need to calculate, first, $W(t)$ which is determined by fluctuations. In order to do this we solve (\ref{phi0s}) and (\ref{cor}) for fluctuations seeking for $\langle \hat\phi^2 \rangle$ in the form $\langle \hat\phi^2 \rangle =   c\cos \omega_0 t + s \sin \omega_0 t$, where $\omega_0 \equiv 2\pi \bar I$, and $\langle \cdots \rangle_t$ denotes averaging in time. We assume also that $c,s \ll 1$. Substituting this form into (\ref{cor}) and keeping only leading terms we obtain in the limit of low temperatures
\begin{equation}
  \langle \hat\phi^2 \rangle = \frac{K_\rho}{2} \left[ \ln\frac{\Lambda}{b} - \frac{\pi W_0}{\omega_0} \cos \omega_0 t\langle \hat\phi^2 \rangle_t - \frac{2W_0}{\omega_0}  \ln\frac{\omega_0}{b} \sin \omega_0 t \right],
  \label{cor2}
\end{equation}
where $W_0= 2W_i K_\rho e^{-2\langle \hat \phi^2 \rangle_t},\; b= \langle W(t) \rangle_t = |c|W_0$. 

Thus we have found that the main logarithmic contribution to $\langle \hat\phi^2 \rangle_t$ is determined by relation similar to (\ref{flu-st}) valid in case of small voltages, but with different infrared cut-off frequency $b$ which is much smaller than $W_c$ in (\ref{flu-st}). From the self-consistency condition we find
\begin{equation}
  c = - \frac{\pi K_\rho W_0}{2\omega_0},  \; s =  \frac{2 c}{\pi} \ln \frac{2\omega_0^2}{\pi W_0^2}, \;  W_0 =  W_i K_\rho^{\frac{1+ K_\rho}{1 - 2 K_\rho}}  \left(\frac{ \pi W_i^2}{2\Lambda V}\right)^{\frac{ K_\rho}{1 - 2 K_\rho}}.
  \label{dV}
\end{equation}
Here we have expressed $W_0$ from (\ref{VT}) in terms of
$V_T$ at zero temperature in the limit of the long wire.

We see that at high voltages the solution with finite
amplitude of the oscillations exists only at $K_\rho < 1/2$,
i.e., when  inter-electronic interaction is strong enough.
The result differs from that for the regime of small
voltages when fluctuations do not destroy the dynamic regime
at any repulsion strength, $K_\rho < 1$. 

Now, using (\ref{cor2}), we can solve (\ref{phi0s}) in the limit of high voltages, $V \gg V_T$, and calculate current. The total current calculating near the contact consists of dc part, $\bar I = V G_0 + I_{nl}$, where $I_{nl}$ is non-linear correction to Ohm's law, and of ac part, $I_{ac} \sin \omega_0 t$, which oscillates with frequency $\omega_0 = 2\pi \bar I/e\approx eV/\hbar$ (in dimensional units)
\begin{eqnarray}
  &&
  I_{ac} = \frac{\sqrt{2} G_0 W_0 }{\sqrt{1+K_\rho^2-(1-K_\rho^2)\cos\omega_0 t_L}}, 
  \label{Iac} \\
  &&  
  I_{nl} = - \frac{2G_0  W_0^2}{V}  \left[\ln \frac{2V^2}{\pi W_0^2} + \frac{1}{(1+ K_\rho^2) - (1 - K_\rho^2)\cos \omega_0 t_L}\right].
  \label{Inl}
\end{eqnarray}
The oscillating factors in these expressions are due to
reflections of current pulses generated at the impurity from
the contacts. The presence of such characteristic oscillations in the static I-V curves was first noted by Dolcini et al ~\cite{Dolcini}.

In the same approximation we can calculate the noise spectrum and we find two maxima of the noise spectrum around frequencies $\omega = \pm \omega_0$
\begin{equation}
  \langle  \delta \hat I(\omega)  \delta \hat I(\omega') \rangle \approx  \frac{\pi  \Gamma (1-2 K_\rho) \sin \pi K_\rho G_0^2 V_T^{2(1-K_\rho)} \delta (\omega + \omega')}{2 (1-K_\rho)^{1-K_\rho} K_\rho^{3K_\rho}||\omega|- \omega_0|^{1-2K_\rho}}   .
  \label{noise}
\end{equation}
Note that the maxima are present under the same condition $K_\rho < 1/2$ for which the solution with finite amplitude of the oscillations at high voltages was found. According to (\ref{noise}) the integral noise power is of the order of $\sim G_0^2 W_i^2$ which is much larger than the ac signal power $ \sim I_{ac}^2$ at frequency $\omega_0$.

In case of long-range Coulomb interaction the correlation function can be found similarly, and we find at $\omega \gg v_F/L$ 
$$
\langle  \delta \hat I(\omega)  \delta \hat I(\omega') \rangle \sim \frac{W_i^2}{8\gamma ||\omega|- \omega_0| \ln\frac{2v_F}{\omega d}}\left(\ln \frac{|\omega|||\omega|- \omega_0|}{W_i^2}\right)^{\frac{1}{4\pi\gamma}} \delta (\omega + \omega').
$$

\subsection{Validity of Gaussian approximation}\label{nonG}

Now we discuss conditions under which the Gaussian model that we have used to describe fluctuations can be justified quantitatively. Note that fluctuations of the displacement field $\hat \phi$ in pure 1D system are Gaussian because the TL Hamiltonian is quadratic, and mean square fluctuations are infinite, $\langle  \hat\phi^2 \rangle = \infty$. Impurity makes fluctuations at the impurity site finite, see (\ref{phi2}), but fluctuations become non-Gaussian because of the cosine impurity term in the Hamiltonian. As the current passes the impurity, the impurity term oscillates, and frequency of the oscillations increases with voltage increasing. This results in a decrease of the time-averaged impurity potential making the impact of the impurity effectively smaller. Therefore one should expect that relative contribution of the non-Gaussian part to fluctuations must decrease in comparison with the Gaussian part. Then at voltages $V \gg V_T$ we can try to calculate non-Gaussian contribution to fluctuations perturbatively. 

We select two contributions of the fluctuating part of the phase, $\hat\phi = \hat\phi_G+\hat\phi_1$, where the first term is the Gaussian contribution which satisfies simplified equation (\ref{phi-flu}), while $\hat\phi$ satisfies full equation (\ref{phi0op}). Considering non-Gaussian part $\hat\phi_1$ as a small correction we linearize (\ref{phi0op}) and obtain equation for $\hat\phi_1$. Considering, again, zero temperature and long conducting channel, $V_T t_L \gg 1$, when $Z(t) \approx K_\rho  \delta (t)$ we derive equation of motion for the third cumulant $C_3$. In the first approximation, $C_3(t) = \langle \hat \phi_1 (t) \hat\phi_G (0)^2 \rangle$. 
$$
\partial_t C_3(t) +  W (t)  C_3 (t) = 4 W_i K_\rho h(t) \sin 2 \Phi (t) \langle \{\phi_G(t)  \phi_G(0)\}\rangle^2 .
$$
Solution of this equation has the form
$$
C_3(t) = -\int_{-\infty}^t dt_1 e^{ -  \int_{t_1}^{t} W  (t_2) dt_2 } 4 W_i K_\rho h(t_1)  \sin 2 \Phi (t_1) \langle \{\phi_G(t_1)  \phi_G(0)\}\rangle^2 
$$
Calculating the integral,  and keeping the leading terms we find
\begin{equation}
  C_3(0) \approx  
  0.35 K_\rho\left[1  - K_\rho \ln \frac{4V^2}{\pi W_0^2}  \right].
  \label{c3}
\end{equation}

Similarly, we can calculate the fourth cumulant  $C_4 = \langle \hat \phi_1  \hat\phi_G (0)^3 \rangle $. Then we can compare non-Gaussian contributions with Gaussian contributions (\ref{cor2}), and find that non-Gaussian contributions are relatively small compared to Gaussian contributions, $C_3 \ll \langle \hat\phi_G^2 \rangle^{3/2}$, $C_4 \ll \langle \hat\phi_G^2 \rangle^{2}$ at small $K_\rho$ and large voltages.

\section{DYNAMIC REGIME OF CONDUCTION IN THE SPINFUL LUTTINGER LIQUID}\label{spinful}

\subsection{Refermionization in the spin channel}\label{Re-spin}

In the spinful LL, similarly to the results of Sec.~\ref{spinless}, the Gaussian approximation for fluctuations in the charge channel 
can be justified in the limits of strong interaction and at high voltages. But in the spin
channel, fluctuations at the impurity site are always non-Gaussian. However, if interaction is spin-rotation
invariant ($K_\sigma =1$) and the impurity is situated in the middle of the wire
we can solve the
problem strictly using the refermionization method. This
method consists in introducing new fermionic variables for
spin channel. Equations of motion for these variables
are linear, and, hence, soluble. Refermionization was
used successfully to treat charge fluctuations in the
spinless case for the specific value of interaction
parameter $K_\rho = 1/2$ \cite{Grabert,
  Chamon_Freed_Wen} and to describe spin fluctuations in the
spinful case for $K_\sigma =1$ \cite{Matveev,ArtVaRe}. Following the
approach of Ref.\cite{Grabert} we introduce new phase
fields 
\begin{equation}
  \hat \phi_{\pm} (x) = \frac{1}{\sqrt{2}}\left[\hat \Phi_{\sigma} (x) + \hat \Theta_{\sigma}(x) \right] \pm \frac{1}{\sqrt{2}}\left[\hat \Phi_{\sigma} (-x) - \hat \Theta_{\sigma}(-x) \right].
  \label{boso}
\end{equation}
New fields are completely decoupled and the impurity term couples to the field $\hat \phi_+$ only. Then we introduce new fermion variables
\begin{equation}
  \sqrt{\frac{1}{2\pi a}}e^{i\hat \phi_+} = \hat g \hat \psi, \quad \hat g = \hat c + \hat c^\dag, \label{e2}
\end{equation}
where $\hat g/\sqrt{2}$ is an auxiliary Majorana
fermion operator, and derive equations of motion for
Heisenberg operators $\hat \psi$ and find that they depend
on $x-v_F t$. Equations of motion for operators  $\hat
\psi_{1,2} (t) = \hat \psi (x= \mp 0, t)$  at the impurity
site and for $\hat g$ have the form
\begin{equation}
  v_F (\hat \psi_2 - \hat \psi_1) = i \hat g f, \quad 
  \partial_t \hat g = i [f(\hat \psi_1 + \hat \psi_2) -  f (\hat \psi^\dag_1+ \hat \psi^\dag_2)],
  \label{dtg}
\end{equation} 
where $f(t) = \sqrt{2\pi a}  W \cos \sqrt{2}\hat \Phi_{\rho}.$

Density perturbations of new fermions are related in a standard way to the gradient of the displacement field
\begin{equation}
  \hat\psi^{+}_{x}\hat\psi_{x} - \langle \hat\psi^{+}_{x}\hat\psi_{x} \rangle_0 = \frac{1}{2\pi}\partial_x\hat\phi_{+}(x).
  \label{ref-density}
\end{equation}

Consider, again, the limit of strong electron-electron interaction when fluctuations in the charge channel are small and represent the field operator at the impurity site 
as a sum of its expectation value and fluctuating part,
$\hat\Phi_\rho  = \Phi_\rho + \hat\phi_\rho$, $\Phi_\rho =
\langle \hat\Phi_\rho\rangle$, taking into account the
fluctuations $\hat\phi_\rho$ in the linear
approximation. Then commutators of $f$ at different times
are small and we can ignore time-ordering and solve
(\ref{dtg}) for $\hat g(t)$ 
$$
\hat g  = 2i \int^t dt_1 [ f(t_1) \hat \psi_1 (t_1)- f(t_1) \hat \psi^\dag_1(t_1)] \exp{\left[-\frac{2}{v_F} \int^t_{t_1} f(t_2)^2 dt_2\right]}.
$$
Now using (\ref{dtg}) and anticommutator $ \{\hat g (t),\hat \psi^\dag_1 (t)\} = \frac{if}{v_F}$ we can obtain the following expression for
$\cos \sqrt{2}\hat \Phi_{\sigma}$: 
\begin{eqnarray}
  &&
  \cos \sqrt{2} \hat \Phi_\sigma (t) = 2 i \pi a W \int^t_{-\infty} \! dt_1  \cos \sqrt{2} \hat \Phi_\rho (t_1) e^{\left[-\frac{2}{v_F} \int^t_{t_1} f(t_2)^2 dt_2\right]}
  \label{I2}
  \\
  &&
  \times  \left\{
    [\hat \psi_1 (t_1)- \hat \psi^\dag_1(t_1)]\hat \psi_1 (t) +  \hat \psi^\dag_1(t)[\hat \psi_1 (t_1)- \hat \psi^\dag_1(t_1)]
  \right\} 
  .
  \nonumber
\end{eqnarray}

We insert (\ref{I2}) into the equation of motion for the
charge phase (\ref{em-rho}). In the limit of small
fluctuations averaging over charge and spin
variables can be performed separately since the fluctuations
in spin and charge sectors are independent. Expectation
values of fermionic densities in averaged  equation (\ref{I2}) can be  associated with distribution function of
new fermions by the relation
\begin{equation}
  \langle \hat \psi^\dag_1 (t_1),\hat \psi_1 (t_2) \rangle = \int \frac{d\varepsilon}{2\pi v_F} n (\varepsilon, t) e^{i\varepsilon (t_1-t_2)}, 
  \label{n}
\end{equation}
where $t =(t_1 +t_2)/2$. Pairings $\langle \hat \psi_1 (t_1),\hat \psi_1 (t_2) \rangle =0$ because operators with subscript 1
are related to the incident spin excitations which are not affected by the impurity because
the coefficient of reflection from the contact
$r=\frac{1-K_\sigma}{1+K_\sigma}$ is equal to zero for
$K_\sigma=1$. Note that this is different from the
case of charge channel considered in Ref.~\cite{Grabert}, because charge excitations incident on
the impurity contain the fraction transmitted through the
impurity and reflected then from the contact.

Now we need to find distribution function $n(\varepsilon, t)$. To do this we, first, subtract boundary conditions (\ref{bc-operator}) at $x=-L/2$ and $x=L/2$ for spin sector and obtain
\begin{equation}
  v_F \partial_x \hat\phi_{+}\left(-\frac{L}{2}, t \right) = \hat P_\sigma .
  \label{bc-refermion}
\end{equation}
We express the derivative  $\partial_x\hat\phi_{+}$ using (\ref{ref-density}), take the expectation value and find the condition for the fermion density expressed in terms of the distribution function
\begin{equation}
  \int \frac{d\varepsilon}{2\pi} [n (\varepsilon, t) - n_F (\varepsilon)] = V_\sigma (t).
  \label{refermion-cond}
\end{equation}
Next we multiply equations (\ref{bc-refermion}) taken at different times $t_1$ and $t_2$ and calculate the expectation value. Reducing products of four fermions to sum of products of pairs in a standard way and using (\ref{P-corr}) we end up with the kinetic equation 
\begin{equation}
  \int
  n (\varepsilon-\omega, t)[1-n (\varepsilon , t)]d\varepsilon=\frac{\omega}{2}
  \left(1+\coth\frac{\omega}{2T}\right). \label{ke}
\end{equation}
Solution of equations (\ref{refermion-cond}-\ref{ke}) has the form of the equilibrium function with the chemical potential equal to spin bias
\begin{equation}
  n (\varepsilon , t) =[1+e^\frac{\varepsilon - V_\sigma (t)}{T}]^{-1}.
  \label{n-sol}
\end{equation}
Note that the distribution function has such a form because at $K_\sigma =1$ there are no reflections of excitations from the contacts. In case of spinless electrons with $K_\rho =1/2$ we would obtain kinetic equation different from (\ref{ke}) which does not have solution in the form of the equilibrium distribution because particles, incident on the impurity, contain a fraction that passed the impurity and then reflected ($r = \frac{1-K_\rho}{1+K_\rho} = \frac{1}{3}$) from the contact. Therefore, the equilibrium form of the distribution function of fermions assumed in Ref.~\cite{Grabert} needs a justification.

Using (\ref{n-sol}) in (\ref{n}) we insert (\ref{I2}) into (\ref{em-rho}) and perform integration over energies. Then we find closed equation for the charge phase
\begin{eqnarray}
  &&
  \partial_t \hat\Phi_{\rho} +  \frac{w}{\sqrt{2}} Z \otimes   \sin \sqrt{2} \hat\Phi_{\rho}(t)  \int^{\infty}_{0} dt_1 \frac{T\cos\sqrt{2} \hat\Phi_{\rho}(t-t_1)}{\sinh \pi T t_1}  
  \label{I4}
  \\
  &&
  \times e^{-2w \int^{t}_{t-t_1} \cos^2\sqrt{2} \hat\Phi_{\rho}(t_2) dt_2} \cos{V_\sigma \left(t-\frac{t_1}{2}\right)t_1} = F \otimes  \hat P_\rho  ,
  \nonumber
\end{eqnarray}
where $w = 2\pi a W^2/v_F$ is the characteristic potential related to the impurity potential renormalized by spin fluctuations. This expression is strict in the limit of strong interaction, and now we will discuss the conditions of validity of our approach that assumes smallness of the fluctuations at the impurity site.

To estimate fluctuations we will simplify (\ref{I4}) taking into account logarithmic divergence of the integral at $t_1 =0$. Then with the logarithmic accuracy we
perform integration neglecting $t_1$ dependence of the regular part of the integrand and using the standard ultraviolet cut-off of the integration at $t_1 \sim 1/\Lambda$. This gives
\begin{equation}
  \partial_t \hat\Phi_{\rho} +  V_0 Z \otimes  \sin 2\sqrt{2} \hat\Phi_{\rho}
  = F \otimes  \hat P_\rho, \quad V_0 = \frac{w}{2\pi\sqrt{2}} \ln\frac{\Lambda}{w},
  \label{em-fluc}
\end{equation}
This equation is similar to (\ref{phi0op}) for a single-mode LL and can be made identical to (\ref{phi0op}) by changing notations. Therefore, for the case of short-range interaction we can use the results of Sec.~\ref{spinless}. Then we find that in the limit of low voltages fluctuations are small provided $K_\rho \ln \frac{\Lambda}{w} \ll 1$, while from (\ref{cor2}-\ref{dV})  we find that  in the limit of large voltages fluctuations are small under condition $K_\rho \ln \frac{\Lambda \omega_0}{w^2} \ll 1$.

In case of long-range interactions we solve \ref{em-fluc} in linear approximation in fluctuating part of the displacement field $\hat \phi_\rho = \hat \Phi_\rho - \Phi_\rho$, $\Phi_\rho = \langle \hat \Phi_\rho \rangle$ and find 
$$
\hat\phi_\rho = \frac{F(\omega) \delta \hat P_\rho}{-i\omega + C }, \quad C = 2\sqrt{2} V_0 \langle Z \otimes     \cos 2\sqrt{2} \Phi_\rho \rangle_t,  
$$
where $\langle\rangle_t$ means time-averaging. To calculate constant $C$ we must solve (\ref{em-fluc}) for expectation value $\Phi_\rho$. Here we will assume that temperature $T$ is low enough, $T \ll V_0$, and limit our estimation by the cases of small and large voltages. 
According to study of the dynamics in Sec.~\ref{spinless} we obtain, again, with logarithmic accuracy
$$
\langle \delta \hat\Phi_\rho^2 \rangle  \approx \sqrt{ \frac{1}{8 \gamma} \ln\left(\frac{v_F \sqrt{\gamma}}{d V_0}\right) }.
$$
In the limit of large voltages the phase increases linearly $2\sqrt{2}\Phi \approx \omega_0 t$ with $\omega_0 = 2\pi f = 2\pi\bar I \approx 2V$, and with the logarithmic accuracy we find
$$
\langle \delta \hat\Phi_\rho^2 \rangle  \approx  \sqrt{ \frac{1}{8 \gamma}  \ln\left(\frac{v_F \omega_0 \sqrt{\gamma}}{d V_0^2}\right)} .
$$
Then we conclude that in case of long-range Coulomb interaction fluctuations of the displacement field at the impurity site in the charge channel are not large at values of parameter $\gamma$ of the order of unity or larger, which is satisfied for typical values of Fermi velocity, confer (\ref{gamma}). 

\subsection{I-V curves and current oscillations}\label{IV-spin}

To calculate current we must solve  (\ref{I4}). In the quasiclassical limit we can neglect fluctuations and substitute $\hat \Phi_\rho$ by its expectation value $\Phi_\rho$. But it is not simple to find the solution analytically, therefore, we restrict our study by limiting cases. 
The simplest case is the regime of current bias. For the  time-independent voltage bias $V_\sigma$ we obtain
\begin{eqnarray}
  &&
  V (t) = \frac{\omega_0}{2} +   \frac{w}{\sqrt{2}}  \int_0^\infty \! d\tilde t \, d t_1 \,   Y(\tilde t)  \sin \frac{\omega_0(t-\tilde t)}{2}  \cos\frac{\omega_0(t-\tilde t - t_1)}{2}
  \label{V}
  \\
  &&
  \times  \cos{V_\sigma t_1}
  \frac{T }{\sinh \pi T t_1}  \exp\left\{-w t_1 - z \;\cos \left[\frac{\omega_0 t_1}{2} + \omega_0( t-\tilde t)\right] \right\} ,
  \nonumber
\end{eqnarray}
where $Y(\omega) = Z(\omega)/F(\omega)$, $z=\frac{2w}{\omega_0}\sin \frac{\omega_0 t_1}{2}$.

\begin{figure}[!ht] 
  \vskip 0mm \centerline{ \psfig{figure=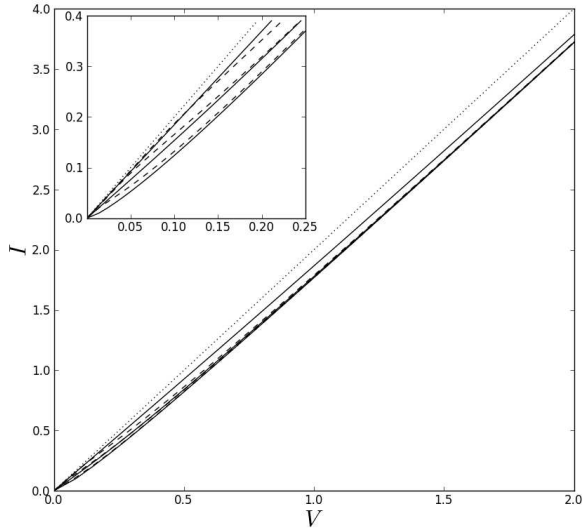,height=8cm
      ,angle=0} }
  \caption{I-V curves at different temperatures $T$ and spin bias
    $V_\sigma$. Voltage is measured in units of the characteristic
    potential $w$, and current in units of $w/G_0$. Dotted line: Ohm's
    law $I=2G_0 V$. Solid lines: $V_\sigma = 0$ for $T = 0, 0.2w,
    1.0w$ from bottom to top.  Dashed lines: $T=0$ for $V_\sigma =
    0.1w, 0.3w, 0.5w$ from bottom to top. The initial part of the I-V
    curves is shown in the inset.} \label{fig-IV}
\end{figure}

If we perform time averaging of (\ref{V}) we find the static I-V curves. 
$$
V_{dc} = \frac{\omega_0}{2} \label{V2}  +    \frac{w}{4} \int^{\infty}_{0} d t_1 \frac{T \cos{V_\sigma t_1}}{\sinh \pi T t_1}
\left[ \sin\omega_0 t_1 \; I_1 \! \left( z \right) + \sin \frac{\omega_0 t_1}{2}\,I_0 \!\left( z \right) \right]e^{-wt_1} ,
$$
This result is the same for both short-range and long-range interaction, which is not very strange since we consider here the limit of strong interaction, and fluctuations in the charge channel are neglected. The general view of the I-V curves for different values of the spin bias is presented in Fig.~\ref{fig-IV}. In the limit of small current, $\omega_0 \ll w$, the second term dominates and $I \propto \sqrt{\frac{\omega_0}{w}}$. 
In the opposite limit of large currents, $\omega_0 \gg w$, the results are similar for both voltage and current bias and the asymptotic I-V curve is parallel to the Ohm's law corresponding to conductance quantum $2G_0 = e^2/(\pi \hbar)$ with the excess voltage $V_{exc} =\frac{w}{8}$.

Time dependence of voltage (\ref{V}) can be characterized by amplitudes of harmonics $n>0$.  
At small currents, $\omega_0 \ll w$, amplitude of harmonics decays slowly, approximately as $1/\sqrt{n}$.
In the limit of large currents, $\omega_0 \gg w$, harmonics decay as power law, and with logarithmic accuracy we obtain for $n>0$
$
V_{n} \approx \frac{w}{8\pi} |Y(n \omega_0)| \left(\frac{w}{2\omega_0}\right)^{n-1}\ln\frac{\Lambda}{\omega_0}.
$

Consider now the case of the voltage bias when the system is driven by external voltage $V + V_1 \cos \omega t$, and
assume the limit of large voltages, $V \approx \omega_0 \gg w$ when the second term in the
left-hand-side of (\ref{I4}) is a small perturbation.
The ac voltage modifies I-V curves, and the most impressive part of this modification is the
resonant steps analogous to the Shapiro
steps in the Josephson junctions. In
contrast to Josephson junctions these steps are not at
constant voltage, but at constant current $I=ef$ like in the
regime of Coulomb blockade~\cite{AverinLi} and in the regime
of sliding CDW in linear-chain conductors~\cite{Gruner}. At
this current the frequency of the ac voltage is equal to the
frequency of current oscillations in the wire. The width of
the step at $V \gg w$ and $V_1 \ll V$ can be calculated
straightforwardly using perturbative approach. We find with
logarithmic accuracy
$$
V_{step}= \frac{V_1 w}{\pi V} |F(\omega_0)| \ln \frac{\Lambda}{\omega_0}.
$$

Non-zero dc spin bias induces a spin current which contains both dc and ac parts. The spin current can be calculated according to relation 
$
I_\sigma = \frac{\sqrt{2}}{\pi}\partial_t \langle \hat \Phi_\sigma \rangle,
$
where $\hat \Phi_\sigma $ can be found from equation of motion (\ref{em-sigma}) using equations (\ref{e2}) and (\ref{n}). 
In the limit of large voltages, $\omega \gg w$ the spin current can be presented in a simple analytic form 
$$
I_\sigma = \frac{V_\sigma}{2\pi}\left(1 + \frac{ w}{\pi \omega_0} \sin \omega_0 t \right).
$$

\section{NON-IDEAL CONTACT}\label{N-ideal}

As non-ideal contacts induce Friedel oscillations in the
quantum wire, one can expect that  such contacts must induce
the effects in transport which are similar to those in the
system with impurity studied in previous sections. This statement is
supported by the results of our letter~\cite{ArtAsSh} where we have studied the spinless LL with
two identical non-adiabatic contacts. However the problem of
transport through non-adiabatic contact to a quantum wire
with a spinful interacting electron gas was not solved. In
this section we consider electronic transport through a
clean quantum wire described as a spinful LL with one ideal
adiabatic and the second non-ideal contact. The main
difficulty in solving this problem is, again, large
fluctuations of the displacement field $\hat\Phi_\sigma$.
And, again, we solve this problem by means of
refermionization in the spin channel.

To study the role of the non-ideal contacts we act similarly to the previous sections, solving equation of motion for the displacement fields with boundary conditions.
Consider boundary conditions for the spin channel with $K_\sigma=1$ with ideal adiabatic contact at $x=L$ and non-adiabatic contact at  $x=0$. Then boundary conditions read
\begin{eqnarray}
  &&
  (v_F \partial_x  - \partial_t )\hat\Phi_\sigma (x=0)  = \hat P^{L}_\sigma -\sqrt{2}f\varepsilon_F \sin \sqrt{2}\hat\Phi_\sigma \cos\sqrt{2}\hat\Phi_\rho \label{bc-1nidal} \\
  &&
  (v_F \partial_x + \partial_t)\hat\Phi_\sigma (x=L)
  = \hat P^{R}_\sigma . \nonumber
\end{eqnarray}

As $\hat\Phi_\sigma (x,t)$ satisfies the equation of
motion
\begin{equation}
  \left(  v_F^2 \partial^2_{x} - \partial^2_{t}\right)\hat\Phi_\sigma (t,x) = 0,
  \label{phiB}
\end{equation}
we can find solution for $\hat\Phi_\sigma (x,t)$ in terms of its values at the contacts. Using then boundary conditions (\ref{bc-1nidal}) we obtain equation of motion for the displacement field at the non-ideal contact 
\begin{equation}
  \partial_t \hat \Phi_{\sigma} +  \sqrt{2}f\varepsilon_F \sin \sqrt{2}\hat \Phi_{\sigma}\cos \sqrt{2}\hat \Phi_{\rho}  
  = \frac{1}{2} [\hat P^{L}_\sigma (t) - \hat P^{R}_\sigma (t-t_L)]  .
  \label{em-sigmaB}
\end{equation}
This equation resembles equation of motion for the displacement field at the impurity site. We map the problem of non-ideal contact to the impurity problem in the LL with ideal contacts at $x= \pm L$ and an impurity characterized by back-scattering matrix element $\tilde W$ at $x=0$. The equation of motion for the displacement field and boundary conditions for such an impurity read
\begin{eqnarray}
  &&
  \left(  v_F^2 \partial^2_{x} - \partial^2_{t}\right)\hat\Phi_\sigma (t,x) =\sqrt{2}v_F   \tilde W \sin \sqrt{2}\hat \Phi_{\sigma}\cos \sqrt{2}\hat \Phi_{\rho} \delta (x), \label{fimp} \\
  &&
  (v_F \partial_x \mp \partial_t)\hat\Phi_\sigma (x= \mp L)
  = \hat Q^{L,R} . \nonumber
\end{eqnarray}
Here we denote external sources of fluctuations as $\hat Q$, and later we will relate them to the source terms $\hat P$. The equation of motion for the phase at the impurity site $x=0$ for this model has a form
\begin{equation}
  \partial_t \hat \Phi_{\sigma} +    \frac{\tilde W}{\sqrt{2}} \sin \sqrt{2}\hat \Phi_{\sigma}\cos \sqrt{2}\hat \Phi_{\rho}
  = \frac{1}{2} [\hat Q^{L} (t-t_L) - \hat Q^{R} (t-t_L)]  .
  \label{em-sigmaFI}
\end{equation}

Comparing now equations (\ref{em-sigmaB}) and (\ref{em-sigmaFI}) we find that equations of motion become identical if we 
choose 
$$
\tilde W = 2 f\varepsilon_F, \quad  \hat Q^{L} (t) = \hat P^{L} (t+t_L), \quad \hat Q^{R} (t) =\hat P^{R}_\sigma (t).
$$

Thus using such substitutions we can use the results for spin channel obtained in Sec~\ref{Re-spin} for quantum wire with one non-ideal contact. 

Now let us consider the charge channel. Following the method used in Sec.~\ref{equations} we find solution of equation of motion for $\hat \Phi_\rho (x,\omega)$ satisfying the boundary conditions. In such a way we obtain expression for the displacement field which depends on the values of both $\hat \Phi_\rho$ and $\hat \Phi_\sigma$ at the boundary with non-adiabatic contact, as both these fields are present in the non-linear term of the boundary condition (\ref{bc-operator-sigma}). Then using this solutions at $x=0$ we find non-linear equations of motion  for $\hat \Phi_\rho (x=0)$ which are similar to (\ref{em-rho}), but with different memory function $Z$ and different right-hand containing the source terms $P_{L,R}$ in a non-symmetric way
$$
\partial_t \hat \Phi_{\rho} +   f\varepsilon_F   Z \otimes  \sin \sqrt{2}\hat\Phi_\sigma \cos\sqrt{2}\hat\Phi_\rho  
=  Z \otimes  \hat P^L_{\rho}  -   F \otimes  \hat P^R_{\rho}.
$$
For the short-range interaction Fourier components of the memory functions read
$$
Z(\omega)= K_{\rho} \frac{1- iK_{\rho} \tan 2\omega t_L}{2 K_{\rho} -i(1+K_\rho^2) \tan 2\omega t_L},  \; F (\omega) =  \frac{K_{\rho} }{2K_{\rho} \cos 2\omega t_L -i(1+K_\rho^2) \sin 2\omega t_L}.
$$
In case of long-range electron-electron interaction we act as in Sec.~\ref{equations} and find similar relations for memory functions but with $K_{\rho}(q_\omega)$ depending on frequency (confer Sec.~\ref{equations}). 

In the limit of strong interaction these equations give the results similar to the case of impurity. 
Thus we find that the problem of electron transport in quantum wire with one non-ideal and the second ideal contacts is mapped to the impurity problem. Then all the results obtained in Sec.~\ref{spinful} can be used for the case of non-ideal contacts if we substitute the impurity potential $W$ for $f\varepsilon_F$ which is the amplitude of the Friedel oscillations induced by the non-ideal contact.

\section{CONCLUSIONS}\label{concl}

Using the approach based on the bosonised Tomonaga-Luttinger
Hamiltonian we have studied electronic transport in 1D
conductors with a single isolated impurity
or with non-ideal contacts to leads of higher dimension, and
predicted a new dynamical regime of conduction in which the
the dc-current is supplemented by ac oscillations with the
wash-board frequency $f = \bar I/e$. 

As thermal fluctuations strongly reduce the effect of impurity on conductance  at temperatures $T > T_0 \sim V_T$, and the effect is also destroyed by fluctuations
in relatively short wires, shorter than the length of the order of $v/V_T$, the dynamic regime predicted in our
work can be observed at low enough temperature in
a relatively long quantum wire, the minimal length and maximal
temperature being related to the magnitude of the defect potential
and the strength of inter-electronic repulsion. 

The impurity potential $W$ can be of different origin and of
different strength. The value of $W$ can be quite small if a
defect is made artificially, say, by a potential of a point
contact. If the defect is induced by an impurity
atom in the conduction channel then the potential can be
quite large, of the order of the Fermi energy.
In semiconductor based quantum wires with shallow impurity
the value of $W$ is expected to be of the order of few millivolts. In
this case the range of frequencies of generated ac signal
can be quite large, up to practically important teraherz
region, depending on material of the quantum wire and the
origin of the defect.

\section*{Acknowledgments}

We are grateful to S. V. Remizov for useful discussions and
helpful comments. The work was supported
by Russian Foundation for Basic Research (RFBR)
and by Fund of non-profit programs ''Dynasty``. 

\appendix
\section{Derivation of the boundary conditions}
Here we derive boundary conditions for a non-ideal contact of the quantum wire with bulky leads.  For the sake of clarity derivation is given for the case of an abrupt rectangular contact but it can be directly generalized using quasiclassical approach for the case of smooth contacts with a sharp potential step.
We start from the expansion for the fermionic field operator over eigen functions of the transverse part of the Hamiltonian in the 1D channel $w_n$ and in the lead $v_n$.

\begin{equation}
  \hat \psi(x,y,z) = \sum\limits_{n=0}^{\infty}
  \hat\psi_n(x)\left[w_n(y,z)\theta(L/2-|x|) + v_n(y,z)\theta(|x|-L/2)
  \right],
  \label{eqn:fermionic-field}
\end{equation}

Then we solve equation of motion for electronic field operators in the
leads using the continuity of both the field operators and their
derivatives at $|x|=L/2$. This allows us to express the solution for
the $n$-th transverse eigenstate in terms of the field operator $\hat
\psi_b$ at the boundary.  Since the results are very similar for both
contacts we concentrate on the left lead. At $x< -L/2$ we obtain a
solution in the left lead
\begin{equation}
  \hat \psi (x) = \hat\psi_b \cos k (x+L/2) + \frac{1}{k} \partial_x \hat\psi_b \sin k (x+L/2),
  \label{psi-operator}
\end{equation}
where $\hat \psi_b$ is the field operator at the left boundary,
i.e. at $x=-L/2$.  This expression contains both incident and outgoing
waves. According to the causality principle, the incident wave $\hat
\psi_{in}(x)$ is determined by a state of the lead far away from the
barrier. Therefore, $\hat \psi_{in}(x)$ must not depend on properties
of the barrier.  Equating the incident part of (\ref{psi-operator}) to
the form describing free particles we find
\begin{equation}
  \hat\psi_b  - \frac{i}{\sqrt{2m(\varepsilon - \varepsilon_n)}} \partial_x \hat\psi_b = \frac{4\pi}{\sqrt{L}} \sum_{k>0} \hat c_{n,s,k} \delta \left(\varepsilon - \varepsilon_n - \frac{k^2}{2m}\right),
  \label{psi-bc}
\end{equation}
where $\hat c_{n,s,k}$ is an annihilation operator of an electron in
the lead.

Equation (\ref{psi-bc}) relates the field operator at the boundary to
the equilibrium states of the $n$-th transverse mode in the lead. We
need a relation between the boundary value of the field operator
corresponding to the lowest transverse eigenstate of the conducting
wire and the incident state of the lead. To find this relation, we
project (\ref{psi-bc}) onto the eigenstates $w_n$ of the wire.

Since
transverse states of the lead are not eigenstates of the wire, we
obtain an infinite system of linear equations for boundary values of
the field operators $\hat \psi_j$ describing different transverse
eigenstates $j$ of the wire
\begin{equation}
  \partial_x\hat\psi_j \delta_{j,0}  + \sum\limits_{j'} r_{jj'} \hat\psi_{j'} = \frac{1}{\sqrt{V}}\sum\limits_{\mathbf{n}=n,k>0}z_{j,n}\hat c_{\mathbf{n}} 2\pi\delta\left(\varepsilon - \varepsilon_{\mathbf{n}} \right),
  \label{psi-bc-system}
\end{equation}
where $z_{j,n} = \langle w_j, v_n \rangle 2ik_{n}$, $r_{j,0} =
\sum\limits_{n} \langle w_j, v_n\rangle ik_{n}\langle v_n, w_0
\rangle$, $r_{j,j'\neq 0} = \sum\limits_{n} \langle w_j, v_n\rangle
ik_{n}\langle v_n, w_{j'} \rangle + \delta_{j,j'} \kappa_j$, and
$k_{n} = \sqrt{2m (\varepsilon - \varepsilon_n^{lead})}$ is a
longitudinal momentum of the $n$-th mode in the lead, $\kappa_j =
\sqrt{2m( \varepsilon_j^{1D} - \varepsilon )}$ is a decay parameter of
the $j$-th mode in 1D, and the Hermitian product is defined in a standard way $\langle f, g \rangle = \int f^*(\mathbf{r}_\bot)g(r_{\bot})d \mathbf{r}_\bot$. The solution of
(\ref{psi-bc})-(\ref{psi-bc-system}) for the lowest subband $j=0$ which
is responsible for an electronic transport in the wire yields the boundary condition for fermionic fields $\hat \psi_0$
\begin{equation}
  A(\varepsilon)\hat\psi_0  + B(\varepsilon) \partial_x \hat\psi_0 =  \frac{1}{\sqrt{V}} \sum_{\mathbf{n}=n,k>0} \gamma_{\mathbf{n}}\hat  c_{\mathbf{n}} 2\pi \delta(\varepsilon - \varepsilon_{\mathbf{n}}),
  \label{psi-bc-0ch},
\end{equation}
where $B = \left\{ r^{-1} \right\}_{0,0} A$, $\gamma_n = A \sum_j \left\{r^{-1}\right\}_{0,j}z_{j,n}$, and
$\left\{r^{-1} \right\}$ is a matrix inverse to $r_{j,j'}$. The
existence of inverse matrix is guaranteed by the existence of the
only solution of matching problem in case of self-conjugate total
Hamiltonian. Note, that coefficients $A$, $B$ and $\gamma$ depend
only on transverse wavefunctions $w_{j}$, $v_n$ and do not depend
neither on the lead at the other end of the 1D channel
nor on the presence of an impurity or electron-electron interaction
if the latter vanishes in the lead including the boundaries. The boundary condition for the right
contact has the same form but with complex-conjugate coefficients.

Although the explicit expressions for the coefficients $A$,$B$ and $\gamma$ are obtained for the case of abrupt
rectangular contacts, the derivation below demands only the linearity of the boundary condition~(\ref{psi-bc-0ch}) and a general requirement of fulfilment of anticommutation relation, and specific values of $A$, $B$ and $\gamma$ are not important.

We can derive a useful relations between coefficients in~(\ref{psi-bc-0ch}) imposing a
requirement of fulfillment of anticommutation relations for electronic
field operators. To do this we consider non-interacting
electrons, for which we can easily solve the equations for the field
operators inside the wire, and calculate the conductance (and, hence,
the transmission probability at the Fermi energy). This allows us to
reduce the number of undetermined constants. We obtain that
anticommutation relations for $\hat \psi_0$ fulfil if and only if
\begin{equation}
  |A^*(\varepsilon)B(\varepsilon)-B^*(\varepsilon)A(\varepsilon)| = \frac{1}{2mV}\sum\limits_{\mathbf{n}=n,k}|\gamma_{\mathbf{n}}|^2 2\pi\delta(\varepsilon-\varepsilon_\mathbf{n}).
  \label{eqn:anticommutator-relation}
\end{equation}
Although we have considered non-interacting electrons in a wire without an impurity,
this relation is valid in general case since $A$, $B$ and $\gamma$ are determined only by transverse wavefunctions.

As it is more convenient to express boundary conditions in terms of
physical values, we multiply (\ref{psi-bc-0ch}) by its Hermitian
conjugate and transform the obtained equation to the time
representation using
relation~(\ref{eqn:anticommutator-relation})
and assuming that the coefficients are slowly varying functions of
energy in the region close to the Fermi energy. Finally, we find the boundary condition~(\ref{bc-rho}) at the left (right) contact for each spin direction. The boundary conditions contain two parameters,
\begin{equation}
  T = \frac{v_F}{V\left( |A(\varepsilon_F)|^2 + q_F^2|B(\varepsilon_F)|^2 \right)}\sum\limits_{\mathbf{n}=n,k}|\gamma_{\mathbf{n}}|^2 2\pi\delta(\varepsilon_F - \varepsilon_{\mathbf{n}}),
  \label{eqn:T}
\end{equation}
$$
f=\frac{V\left(|A(\varepsilon)|^2 - q_F^2|B(\varepsilon)|^2\right)}{\sum\limits_{\mathbf{n}=n,k}|\gamma_{\mathbf{n}}|^2 2\pi\delta(\varepsilon_F - \varepsilon_{\mathbf{n}})}.
$$

To understand the physical meaning of $T$ it is instructive to
consider a system of non-interacting electrons without an impurity in
the wire. Since the problem without the interaction and an impurity can be solved exactly, we can find
a conductance of non-interacting system using equations of motions for non-interacting electrons and boundary conditions~(\ref{psi-bc-0ch}). Then we obtain the conductance $G=G_0 T$. Comparing the latter expression with the Landauer formula
we conclude that $T$ coincides with transmission probability of a non-interacting system without
an impurity, and $0< T\le 1$. Since the right-hand side of~(\ref{eqn:T}) depends only on the properties of the contact the latter
conclusion is valid for a general case
(however, in the general case we cannot claim that $T$ given by~(\ref{eqn:T}) coincides with the transmission probability).

\bibliographystyle{model1a-num-names}
\bibliography{biblio-paper.bib}

\end{document}